\documentclass{ws-ijgmmp}
\usepackage{graphicx}

\begin{document}

\markboth{Thiago Prud\^encio, Diego Julio Cirilo-Lombardo}
{Entanglement and non-trivial topologies}

%
\catchline{}{}{}{}{}
%

\title{ENTANGLEMENT AND NON-TRIVIAL TOPOLOGIES}

\author{THIAGO PRUD\^ENCIO}

\address{Institute of Physics, University of Brasilia -UnB, CP: 04455,
70919-970, Brasilia-DF, Brazil.\\
\email{Corresponding e-mail: thprudencio@gmail.com}}

\author{DIEGO JULIO CIRILO-LOMBARDO}

\address{International Institute of Physics, Universidade Federal do Rio Grande do
Norte, 1722, 59078-400, Natal-RN, Brazil.\\
Bogoliubov Laboratory of Theoretical Physics Joint Institute for Nuclear Research 141980, Dubna-Moscow, Russia.\\
\email{Corresponding e-mail:diego777jcl@gmail.com}}

\maketitle


\begin{abstract}
We consider states defined on non-trivial topologies of Torus,
Mobius and Torus-Mobius. Adequate operators leading to the construction of coherent states, two-mode coherent states 
and entangled states are derived and we show a class of entangled states on Torus that are related 
to the entanglement of photons by orbital angular momentum states. 
\end{abstract}


\section{Introduction}

States in manifolds with non-trivial topologies and entanglement involving such states have 
close implications to quantum information \cite{almeida,ladd,gotesman,wooters,dieks,thiago,thiago2} and quantum computing \cite{barz,golovach,shi,abanto,neegard}. 
More specifically, generation of coherent states and their superpositions (cat states) \cite{schrodinger,huyet} or 
and ther entanglement (as EPR states \cite{einstein}) in intricate 
geometrical or topological configurations can give rise to interesting effects. 
Recent developments of topological qubits \cite{fu,preskill}, 
quantum optics in curved space-time \cite{milburn}, topological defects \cite{zurek}, 
graphene and topological isulators \cite{sun,lee,song,wang} 
have increased the interest in systems with non-trivial topologies. 
Here we concentrate on Torus and M\"obius topologies. 

Different from the known coherent state elaborations for quantum optics \cite{glauber,kim,bayfield,raimond}, 
in non-trivial manifolds, coherent states acquire topological properties, 
with consequences that affect periodicity and normalization. 
This is the case of non-trivial manifolds with topologies of Torus 
\cite{kowalski,kowalski2} and Mobius strip \cite{diego}. 

We remark that there is no general method for a construction of coherent states for
a particle on an arbitrary manifold, specially when the space is described in a non-trivial topology 
\cite{kastrup,cirilo,provost,gazeau2,klauder,barut,gazeau,olmo,kempf}. 
A general algorithm was introduced by Perelomov \cite{peremolov} for construction of coherent
states for homogeneous spaces $X$ which are quotients $X = G/H$ of a Lie-group
manifold $G$ and the stability subgroup $H$. Unfortunately, in many interesting cases
for physics as a particle on a circle,
sphere or torus, the phase space whose points should label the coherent
states, more precisely a cotangent bundle $T*M$, where $M$ is the configuration
space, is not a homogeneous space. A general theory of coherent states when the configuration space 
has non-trivial topology is far from complete. In view of the lack of the general method
for the construction of coherent states one is forced to study each case of
a particle on a concrete manifold separately.

It is worth to point out that coherent states for a quantum particle on circle, sphere and torus have been
introduced very recently \cite{kowalski}. In such cases
different constructions of the coherent states (CS) in the boson case are practically straightforward, and 
a simple addition by hand of $1/2$ to the angular momentum operator $J$ for the fermionic case into the
corresponding CS remains obscure and non-natural. The question that naturally arises is if
there exists any geometry for the phase space in which the CS construction leads precisely
to a fermionic quantization condition. In a recent work \cite{diego}, we demonstrated the positive
answer to this question showing that the CS for a quantum particle on the Mobius strip geometry 
is the natural candidate to describe fermions exactly as the cylinder geometry for
bosons. 

In this paper, we investigate the entanglement of coherent states on a Torus, Mobius strip and 
Torus-Mobius, i.e., in the case where two-particle states, one in the Torus and other in Mobius, are entangled in the 
intersection of the two manifolds. Taking into account the consistent operators acting in each space and their 
action on states, we consider the generation of entangled states and non-orthogonal projective measurements, a generalization 
of von-Neumann type measurement, to quantify 
entanglement. 
We also consider the topological and geometrical consequences associated to CS in Torus and Mobius and the 
relations of the toroidal operators and entangled torus states to $SU\left( 1,1\right)$. 

\section{States on a torus and operators $\hat{\mathcal{M}}$  }

We can start with a revision of the quantum mechanics on a Torus. Such situation, in principle, can be identified with product of two
circles, this implies, following \cite{kowalski}, that we can define the algebra 
\begin{eqnarray}
\left[J_{i},U_{j}\right]&=&\delta_{ij}U_{j}, \\
\left[J_{i},U_{j}^{\dagger}\right]&=& -\delta_{ij}U_{j}^{\dagger}, \\
\left[J_{i},J_{j}\right]&=& 0,   \\
\left[U_{i},U_{j}\right] &=&\left[U_{i}^{\dagger},U_{j}^{\dagger}\right]= 
\left[U_{i},U_{j}^{\dagger}\right]=0,
\end{eqnarray}
where $U_{i}=e^{i\hat{\phi}_{i}}$ and $U_{i}^{\dagger}=e^{-i\hat{\phi}_{i}}$. 
Taking $i,j=1,2$, and $\vec{J}=(J_{1},J_{2})$, $\vec{j}=(j_{1},j_{2})$,
the eigenvalue equation is written as 
\begin{eqnarray}
\vec{J}|\vec{j}\rangle=\vec{j}|\vec{j}\rangle.
\end{eqnarray}
where $|\vec{j}\rangle$ are ladder operators that satisfy 
\begin{eqnarray}
e^{i\hat{\phi}_{j}}|\vec{j}\rangle &=& |\vec{j}+\vec{e}_{j}\rangle,  \notag
\\
e^{-i\hat{\phi}_{j}}|\vec{j}\rangle &=& |\vec{j}-\vec{e}_{j}\rangle,
\end{eqnarray}
where $\vec{e}_{1}=(1,0)$ and $\vec{e}_{2}=(0,1)$. Consequently, starting
from a vector $|\vec{j}_{0}\rangle$, and $j_{0i}\in [0,1)$, $i=1,2$, a
Hilbert space can be generated by a set of vectors $\lbrace |\vec{j}\rangle
\rbrace$. The antiunitary operator of time inversion $T$ can be also defined
by means of the relations 
\begin{eqnarray}
TJ_{i}T^{-1}&=&-J_{i} \\
Te^{i\hat{\phi}_{i}}T^{-1}&=& e^{-i\hat{\phi}_{i}} \\
Te^{-i\hat{\phi}_{i}}T^{-1}&=& e^{i\hat{\phi}_{i}}
\end{eqnarray}
This operator is symmetric and the action on $|\vec{j}\rangle$ is given by 
\begin{eqnarray}
T|\vec{j}\rangle=|-\vec{j}\rangle.
\end{eqnarray}
As a consequence $\vec{j}_{0}$ can take only four possible values $(0,0)$, $%
(0,\frac{1}{2})$, $(\frac{1}{2},0)$ and $(\frac{1}{2},\frac{1}{2})$.
The relation between $|\vec{j}\rangle $ and $|\vec{\phi}\rangle $ is given
by 
\begin{equation}
\langle \vec{\phi}|\vec{j}\rangle =e^{i\vec{j}\cdot \vec{\phi}} 
\end{equation}
In the coordinate representation of the states on a torus, the following
eigenvalue equation is satisfied 
\begin{equation}
e^{\pm i\hat{\phi}_{i}}|\vec{\phi}\rangle =e^{\pm i\phi _{i}}|\vec{\phi}%
\rangle , 
\end{equation}%
where 
\begin{equation}
\int_{0}^{2\pi }\int_{0}^{2\pi }\frac{d\phi _{1}d\phi _{2}}{(2\pi )^{2}}|%
\vec{\phi}\rangle \langle \vec{\phi}|=I. 
\end{equation}%

Now, let us define the following operator $\hat{\mathcal{M}}_{i}$ that mixes
operators $e^{i\hat{\phi}_{i}}$ and $T$, 
\begin{equation}
\hat{\mathcal{M}}_{i}=\left( e^{i\hat{\phi}_{i}}T+\left( e^{i\hat{\phi}%
_{i}}T\right) ^{-1}\right) . 
\end{equation}%
The action of this operator on $|\vec{j}\rangle $ will generate
superposition states of the form 
\begin{equation}
\hat{\mathcal{M}}_{i}|\vec{j}\rangle =|-\vec{j}+\vec{e}_{i}\rangle +|-\vec{j}%
-\vec{e}_{i}\rangle 
\end{equation}%
\begin{equation}
\hat{\mathcal{M}}_{i}^{2}|\vec{j}\rangle =2|\vec{j}\rangle , 
\end{equation}%
As a consequence the even exponents of $\hat{\mathcal{M}}_{i}$ are
eigenvectors of $|\vec{j}\rangle $, 
\begin{eqnarray}
\hat{\mathcal{M}}_{i}^{2n}|\vec{j}\rangle &=&2^{n}|\vec{j}\rangle \\
\hat{\mathcal{M}}_{i}^{2n+1}|\vec{j}\rangle &=&2^{n}\left( |-\vec{j}+\vec{e}%
_{i}\rangle +|-\vec{j}-\vec{e}_{i}\rangle \right)
\end{eqnarray}%
We have then
\begin{eqnarray}
e^{\hat{\mathcal{M}}_{i}}|\vec{j}\rangle &=&\cosh (2)|\vec{j}\rangle +\sinh (2)\left( |-\vec{j}+\vec{e}_{i}\rangle +|-\vec{j}-\vec{e}%
_{i}\rangle \right) .
\end{eqnarray}

\section{Relation of the toroidal operators $\hat{\mathcal{M}}~$and $SU\left( 1,1\right)$}

We have been seen in previous references, the action of the creation and
annihilation operators corresponding to the generators of the
Heisenberg-Weyl algebra with the basic coherent states of the non-compact
oscillator is given as follows
\begin{eqnarray}
a^{2n+1}\left\vert +\right\rangle &=&\alpha ^{2n+1}\left\vert -\right\rangle
\\
a^{2n}\left\vert +\right\rangle &=&\alpha ^{2n}\left\vert +\right\rangle
\end{eqnarray}%
and $e^{a}|+\rangle =\cosh (\alpha )|+\rangle +\sinh (\alpha )|-\rangle$,
with $\alpha $ the corresponding eigenvalue of the coherent state.

The importance of such construction is because it describes a unitary
representation of the $SL(2R)\sim SU(1,1)\sim SO(2,1)$ group that is of
infinite dimension. This representattion was used in the case of coherent
states in noncompact oscilatrs and also in Supergravity models and
strings.This specific action of the operators $a^{2n}\left( a^{2n+1}\right) $%
over the basic states $\left\vert +\right\rangle \left( \left\vert
-\right\rangle \right) $ is clearly isomorphous with the action of the$\hat{%
\mathcal{M}}_{i}$ operators defined on the torus making the identification%
\begin{eqnarray}
|\vec{j}\rangle &\rightarrow &\left\vert +\right\rangle, \\
\left( |-\vec{j}+\vec{e}_{i}\rangle +|-\vec{j}-\vec{e}_{i}\rangle \right)
&\rightarrow &\left\vert -\right\rangle,
\end{eqnarray}%
and also 2$\rightarrow \alpha .$ This operator $\mathcal{M}$ make a cross over from a
compact to a non compact structure of the of the geometry of the
configuration space. This point is extreamely important to explain
topologically and geometrically speaking the metal/superconductor-insulator
transition. There exists, for instance, a clear possibility to map the SUGRA (Supergravity)
model \cite{diegoplb} into the Torus due the specific action of the operators $\mathcal{M}$. 
A concrete adaptation of the model to metal isulator-transition and SUGRA-Torus correspondence 
will be the scope of a future work \cite{tdiego}.

In particular, let us first propose an entangled state based on such states
\begin{equation}
|+\rangle |-\rangle +|-\rangle |+\rangle, 
\end{equation}
we will show that there is an explicit operator capable of realize the building of 
such state.

Here we can only analyse the following properties. The change under the action of $ab$, where $b$ has the same action of
the operator $a$, is the following 
\begin{equation*}
ab\left( |+\rangle _{a}|-\rangle _{b}+|-\rangle _{a}|+\rangle _{b}\right),
=\left( |-\rangle _{a}|+\rangle _{b}+|+\rangle _{a}|-\rangle _{b}\right). 
\end{equation*}%
Then, the state is invariant under $ab$. We can also check the action
under $ab^{2}$ and $a^{2}b$, 
\begin{equation*}
ab^{2}\left( |+\rangle _{a}|-\rangle _{b}+|-\rangle _{a}|+\rangle
_{b}\right) =\left( |-\rangle _{a}|-\rangle _{b}+|+\rangle _{a}|+\rangle
_{b}\right), 
\end{equation*}%
\begin{equation*}
a^{2}b\left( |+\rangle _{a}|-\rangle _{b}+|-\rangle _{a}|+\rangle
_{b}\right) =\left( |+\rangle _{a}|+\rangle _{b}+|-\rangle _{a}|-\rangle
_{b}\right). 
\end{equation*}
Such that this state generate two equivalent and entangled states. 

\section{Coherent states on Torus and Mobius}

The coherent states on a Torus are defined from the relation 
$\vec{Z}|\vec{z}\rangle =\vec{z}|\vec{z}\rangle$, 
where $\vec{z}=(z_{1},z_{2})\in \mathbb{C}^{2}$, $\vec{Z}=(Z_{1},Z_{2})$ and 
$Z_{i}=e^{-J_{i}+\frac{1}{2}}e^{i\hat{\phi}_{i}}, z_{j}^{(Torus)} =e^{-l_{j}+i\alpha _{j}}$.
The relation to the states $|\vec{j}\rangle $ is given by 
\begin{equation}
\langle \vec{j}|\vec{z}\rangle =z_{1}^{-j_{1}}z_{2}^{-j_{2}}e^{-\vec{j}%
^{2}/2}. 
\end{equation}
The coherent states can be explicitly written in terms of $|\vec{j}%
\rangle $, using $\sum_{\vec{j}\in \mathbb{Z}^{2}}|\vec{j}\rangle \langle \vec{j}|=I$, by means of 
\begin{eqnarray}
|\vec{z}\rangle^{(Torus)} &=&\sum_{\vec{j}\in \mathbb{Z}%
^{2}}z_{1}^{-j_{1}}z_{2}^{-j_{2}}e^{-\vec{j}^{2}/2}|\vec{j}\rangle, \label{z}
\end{eqnarray}%
that can also be written as
\begin{eqnarray}
|\vec{z}\rangle^{(Torus)} =\sum_{\vec{j}\in \mathbb{Z}^{2}}e^{\vec{l}\cdot \vec{j}-i\vec{\alpha}%
\cdot \vec{j}}e^{-\vec{j}^{2}/2}|\vec{j}\rangle. 
\end{eqnarray}
The normalization conditions of (\ref{z}) are product of Jacobi theta
functions, here we consider only the case $(0,0)$, that is given by 
\begin{equation*}
\langle \vec{z}|\vec{z^{\prime }}\rangle =\Theta _{3}\left( \frac{i}{2\pi }%
\ln (z_{1}^{\ast }z_{1}^{\prime })|\frac{i}{2\pi }\right) \Theta _{3}\left( 
\frac{i}{\pi }\ln (z_{2}^{\ast }z_{2}^{\prime })|\frac{i}{\pi }\right) . 
\end{equation*}%
Now, we define two-mode coherent states by means of the product state 
\begin{eqnarray}
|\vec{z},\vec{w}\rangle &=&\sum_{\vec{j},\vec{j^{\prime }}\in \mathbb{Z}%
^{2}}z_{1}^{-j_{1}}z_{2}^{-j_{2}}w_{1}^{-j_{1}^{\prime
}}w_{2}^{-j_{2}^{\prime }}e^{-\vec{j}^{2}/2}e^{-\vec{j^{\prime }}^{2}/2}|%
\vec{j},\vec{j^{\prime }}\rangle ,  \notag  \label{zz} \\
&&
\end{eqnarray}%
where $w_{j}=e^{-l_{j}^{\prime }+i\alpha _{j}^{\prime }}$. We can also write this state as 
\begin{eqnarray}
|\vec{z},\vec{w}\rangle &=&\sum_{\vec{j},\vec{j^{\prime }}\in \mathbb{Z}%
^{2}}e^{\vec{l}\cdot \vec{j}+\vec{l^{\prime }}\cdot \vec{j^{\prime }}-i\vec{%
\alpha}\cdot \vec{j}-i\vec{\alpha ^{\prime }}\cdot \vec{j^{\prime }}}e^{-(%
\vec{j}^{2}+\vec{j^{\prime }}^{2})/2}|\vec{j},\vec{j^{\prime }}\rangle,
\notag 
\end{eqnarray}
where the relations the eigenvalue relations are written as 
\begin{eqnarray}
Z_{i}|\vec{z},\vec{w}\rangle &=&z_{i}|\vec{z},\vec{w}\rangle , \\
W_{i}|\vec{z},\vec{w}\rangle &=&w_{i}|\vec{z},\vec{w}\rangle ,
\end{eqnarray}%
$W_{i}=e^{-J_{i}^{\prime }+\frac{1}{2}}e^{i\hat{\phi}_{i}^{\prime }}$ and the normalization factor 
\begin{eqnarray}
\langle \vec{z},\vec{w}|\vec{z^{\prime }},\vec{w^{\prime }}\rangle &=&\Theta
_{3}\left( \frac{i}{2\pi }\ln (z_{1}^{\ast }z_{1}^{\prime })|\frac{i}{\pi }%
\right) \nonumber \\
&\times &\Theta _{3}\left( \frac{i}{2\pi }\ln (z_{2}^{\ast }z_{2}^{\prime })|%
\frac{i}{\pi }\right)  \notag \\
&\times &\Theta _{3}\left( \frac{i}{2\pi }\ln (w_{1}^{\ast }w_{1}^{\prime })|%
\frac{i}{\pi }\right) \nonumber \\
&\times& \Theta _{3}\left( \frac{i}{2\pi }\ln (w_{2}^{\ast
}w_{2}^{\prime })|\frac{i}{\pi }\right). 
\end{eqnarray}%
The action of $\hat{\mathcal{M}}_{i}$ on a two-mode coherent state for
operators $\hat{\mathcal{M}}_{i}^{z}$ and $\hat{\mathcal{M}}_{i}^{w}$ acting
in each mode, we will have the following relations for even exponents 
\begin{eqnarray}
\left( \hat{\mathcal{M}}_{i}^{z}\right) ^{2n}|\vec{z},\vec{w}\rangle
&=&2^{n}|\vec{z},\vec{w}\rangle , \\
\left( \hat{\mathcal{M}}_{i}^{w}\right) ^{2n}|\vec{z},\vec{w}\rangle
&=&2^{n}|\vec{z},\vec{w}\rangle .
\end{eqnarray}%
It follows that the two-mode coherent states from the torus are eigenstates of the even
exponents of $\hat{\mathcal{M}}_{i}^{z}$ and $\hat{\mathcal{M}}_{i}^{w}$.

The coherent states can be reduced from the Torus to Mobius strip
 by means of a constraint between the angle variables
\begin{equation}
\theta =\frac{\phi +\pi }{2}.  
\end{equation}
we will describe this more appropriatelly in a section below. Once we have the a Mobius strip parametrization, a 
deformation can be applied by means of following transformation 
\begin{eqnarray}
X^{'(Mobius)} &=&e^{-\mathcal{Z}}X^{(Mobius)},  \label{a1} \\
Y^{'(Mobius)} &=&e^{-\mathcal{Z}}Y^{(Mobius)},  \label{a2} \\
Z^{'(Mobius)} &=& Z^{(Mobius)},  \label{a3}
\end{eqnarray}%
that does not change the topological properties of the Mobius strip.
In this way a coherent state associated to the Mobius strip \cite{diego} is defined by means
of $
\vec{Z}^{(Mobius)}|\vec{z}\rangle =\vec{z}^{(Mobius)}|\vec{z}\rangle$,
where $\vec{z}^{(Mobius)}=(z_{1}^{(Mobius)},z_{2}^{(Mobius)})\in \mathbb{C}^{2}$. $Z_{i}^{(Mobius)}$ acts
in the $i$ mode of $|\vec{z}\rangle$ and 
\begin{equation}
z_{i}^{(Mobius)}=e^{-\left( l_{i}+r\sin \left( \phi _{i}/2\right) \right) +i\phi
_{i}}\left( 1+r\cos \left( \phi _{i}/2\right) \right) .  \label{23}
\end{equation}%

\section{Entangled states on Torus}

Let us consider the action of the following operators 
\begin{equation*}
\hat{D}_{ik}^{n}=\left(\hat{\mathcal{M}}_{i}^{z}\right)^{2n}\left(\hat{\mathcal{M}}_{k}^{w}\right)^{2n+1 }+\left(\hat{%
\mathcal{M}}_{i}^{w}\right)^{2n}\left(\hat{\mathcal{M}}_{k}^{z}\right)^{2n+1} 
\end{equation*} %
on the states $|\vec{j},\vec{j^{\prime }}\rangle $, where $n=0,1,...$ is are integer values,. This will be given by
\begin{eqnarray}
\hat{D}_{ik}^{n}|\vec{j},\vec{j^{\prime }}\rangle &=&2^{2n}|\vec{j}\rangle
\left( |-\vec{j^{\prime }}+\vec{e^{\prime }}_{k}\rangle +|-\vec{j^{\prime }}-%
\vec{e^{\prime }}_{k}\rangle \right)  \notag \\
&+&2^{2n}\left( |-\vec{j}+\vec{e}_{i}\rangle +|-\vec{j}-\vec{e}_{i}\rangle
\right) |\vec{j^{\prime }}\rangle,  \label{dos}
\end{eqnarray}
Note that this state is equivalent to the state $|+\rangle |-\rangle +|-\rangle |+\rangle$,
discussed previously. This state is built in a consistent way from the state
$|\vec{j},\vec{j^{\prime }}\rangle$ by means of the action of the 
operator $\hat{D}_{ik}^{n}$. In particular, the operators $a$ and $b$ discussed previously
are in fact actions of the operator $\mathcal{M}_{i}$ that in appropriate conditions leave the 
state invariant, i.e., in the forms $a^{2}b$ and $ab^{2}$. Such operators are important when 
we search operators that leave a bipartite entangled state invariant \cite{jimenes}.

Now, let us consider the bipartite coherent state for a torus,
\small 
\begin{eqnarray}
\hat{D}_{ik}^{n}|\vec{z},\vec{w}\rangle &=&\sum_{\vec{j},\vec{j^{\prime }}%
\in \mathbb{Z}^{2}}e^{\vec{l}\cdot \vec{j}+\vec{l^{\prime }}\cdot \vec{%
j^{\prime }}-i\vec{\alpha}\cdot \vec{j}-i\vec{\alpha ^{\prime }}\cdot \vec{%
j^{\prime }}}e^{-(\vec{j}^{2}+\vec{j^{\prime }}^{2})/2}\hat{D}_{ik}^{n}|\vec{%
j},\vec{j^{\prime }}\rangle ,  \notag \\
&&
\end{eqnarray}
\normalsize%
this can be written as 
\begin{eqnarray}
\hat{D}^{n}_{ik}|\vec{z},\vec{w}\rangle &=& 2^{2n}\sum_{\vec{j},\vec{j'}\in \mathbb{Z}^{2}} e^{\vec{l}\cdot\vec{j}+\vec{l'}\cdot\vec{j'}-i\vec{\alpha}\cdot\vec{j}-i\vec{\alpha'}\cdot\vec{j'}}e^{-(\vec{j}^{2}+\vec{j'}^{2})/2}|\vec{j}\rangle\left(|-\vec{j'}+\vec{e'}_{k}\rangle 
+ |-\vec{j'}-\vec{e'}_{k}\rangle\right) \nonumber \\
&+& 2^{2n}\sum_{\vec{j},\vec{j'}\in \mathbb{Z}^{2}} e^{\vec{l}\cdot\vec{j}+\vec{l'}\cdot\vec{j'}-i\vec{\alpha}\cdot\vec{j}-i\vec{\alpha'}\cdot\vec{j'}}e^{-(\vec{j}^{2}+\vec{j'}^{2})/2}\left(|-\vec{j}+\vec{e}_{i}\rangle + |-\vec{j}-\vec{e}_{i}\rangle\right)|\vec{j'}\rangle.
\label{eqlerA}
\end{eqnarray}
The case where $|-\vec{j^{\prime }}\pm \vec{e^{\prime }}_{k}\rangle =|-\vec{%
j^{\prime }}\rangle \pm |\vec{e^{\prime }}_{k}\rangle $, we can write 
\begin{eqnarray}
\hat{D}_{ik}^{n}|\vec{j},\vec{j^{\prime }}\rangle =2^{2n+1}\left( |\vec{j}%
\rangle |-\vec{j^{\prime }}\rangle +|-\vec{j}\rangle |\vec{j^{\prime }}%
\rangle \right) . 
\label{eq1er}
\end{eqnarray}%
It is interesting to observe that the states $|-\vec{j}\rangle $ and $-|\vec{%
j}\rangle $ are not equivalents, since $\vec{J}|-\vec{j}\rangle =-\vec{j}|-%
\vec{j}\rangle $ and $\vec{J}(-|\vec{j}\rangle )=\vec{j}(-|\vec{j}\rangle )$%
, what implies that the above state is in fact a entangled state.
\begin{figure}[]
\centering
\includegraphics[scale=0.4]{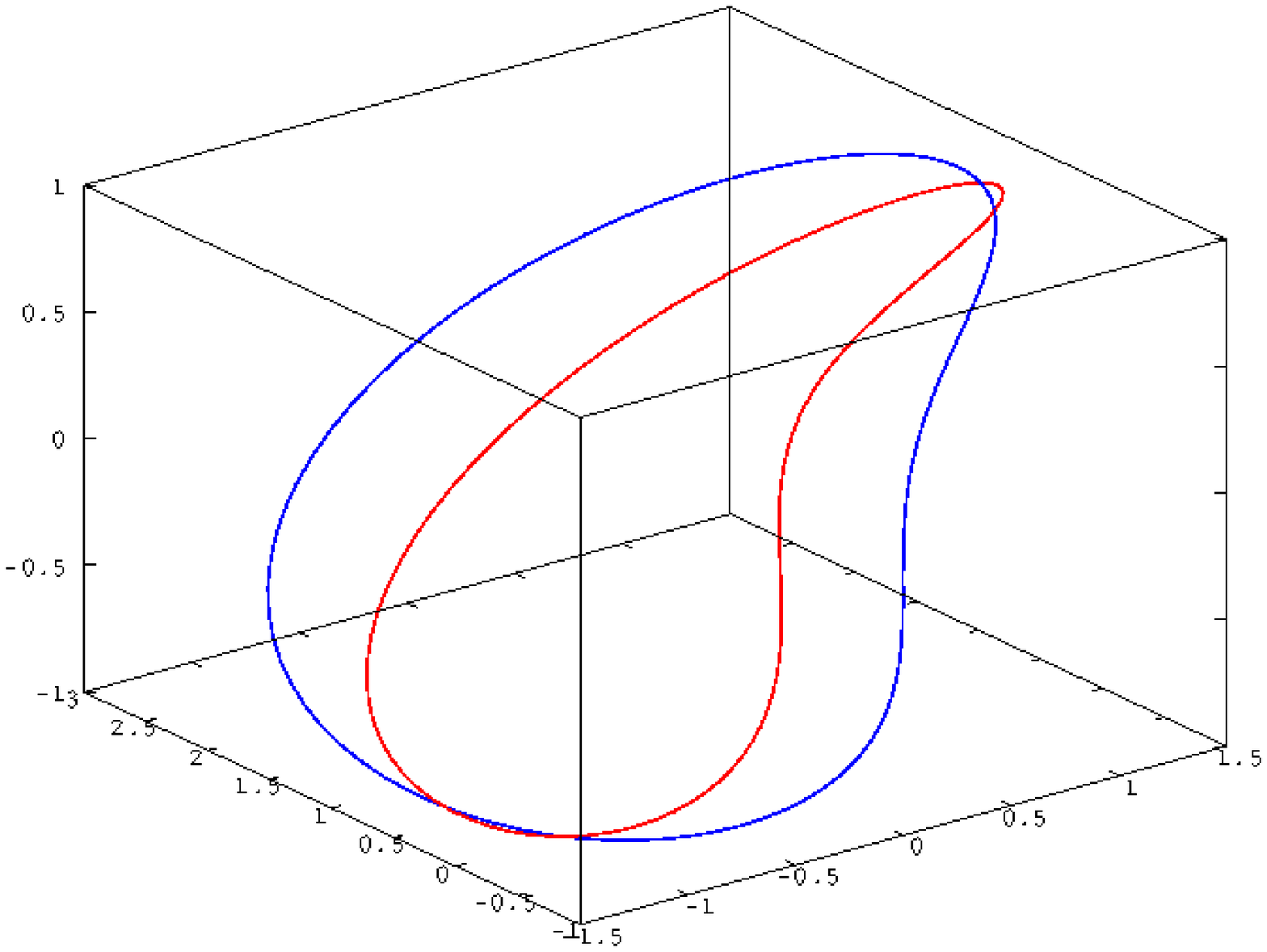}
\caption{(Color online) two-particle trajectories on Torus.}
\label{trajtorus2}
\end{figure}
\begin{figure}[]
\centering
\includegraphics[scale=0.4]{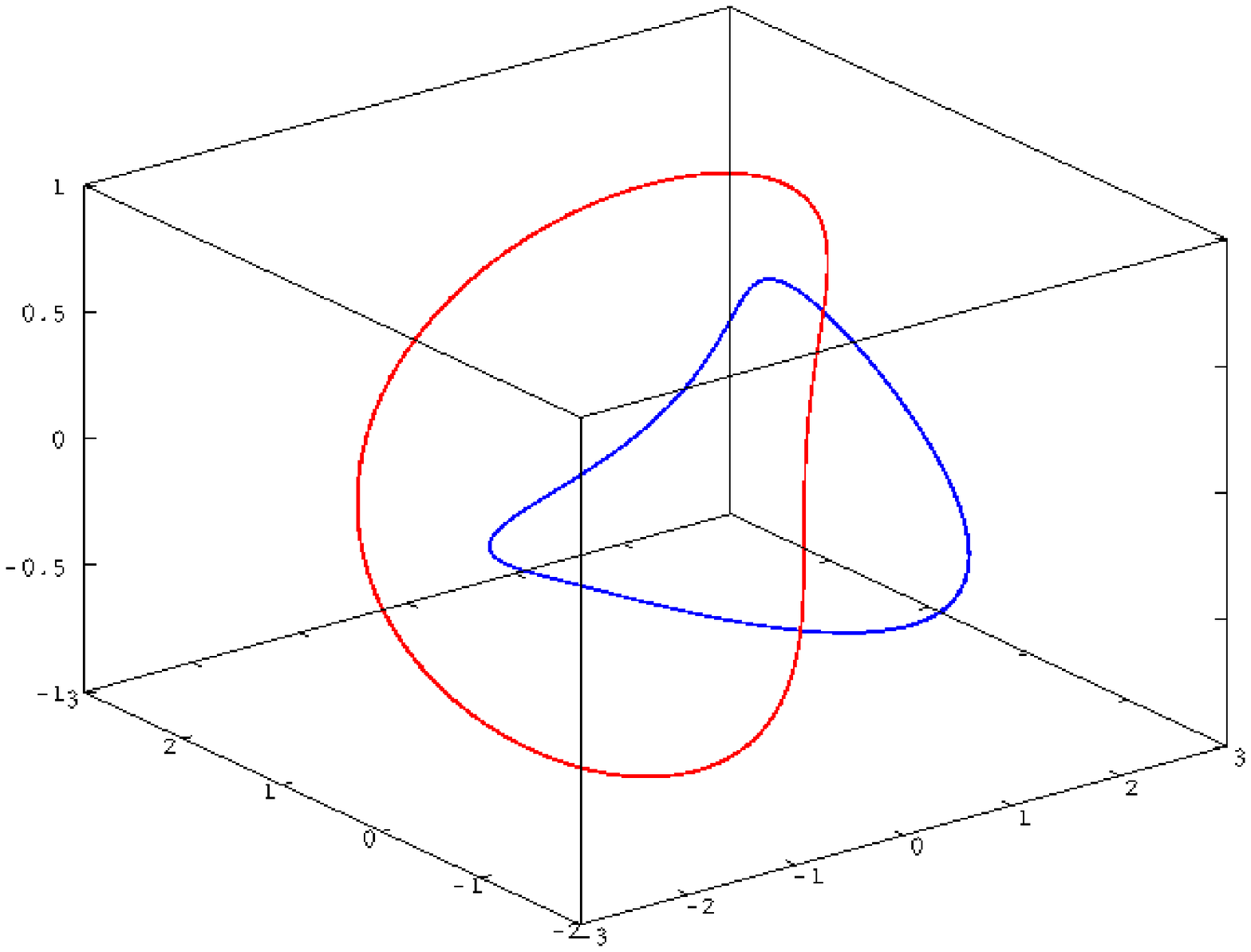}
\caption{(Color online) two-particle trajectories on Torus.}
\label{trajtorus21}
\end{figure} 
Note that the state (\ref{eqlerA}) with (\ref{eq1er}) can be associated to the state of photons with
 orbital angular momentum states $|l\rangle$ entangled, experimentally verified \cite{pors,walborn,vaziri}, described by
\begin{eqnarray}
|\Psi\rangle =\sum_{l}\sqrt{\lambda_{l}}|l\rangle|-l\rangle,
\end{eqnarray}
and $\langle \phi|l\rangle = e^{il\phi}/\sqrt{2\pi}$ where $\phi$ is the azimuthal angle.

In the general case the operator $\hat{D}^{n}_{ik}$ acts in the entanglement of two-particle states 
with torus topology (figure \ref{trajtorus2} and \ref{trajtorus21}).

\section{Geometrical and topological reduction from Torus to Mobius}

In this section we arise to the question about if the topology of a manifold
is sufficient to give a correct description of the physical states living
there. In the reference \cite{diego}, we show the reduction of the toroidal geometry to the Mobius strip due 
the use of suitable projection operators, starting from the torus as the original quantum phase space.

A position point in a
Mobius strip geometry can be parameterized by means of specific points $P_{0}$ and $P_{1}$ given by 
\begin{eqnarray}
P_{0}&=&\left( X_{0},Y_{0},Z_{0}\right) \\
P_{1}&=&\left(X_{0}+X_{1},Y_{0}+Y_{1},Z_{0}+Z_{1}\right)  \label{1}
\end{eqnarray}
where the coordinates of $P_{0}$ describe a central cylinder, generated from a
invariant fiber in the middle of the strip weight, i.e.,
\begin{eqnarray}
X_{0}&=& R\cos\varphi \\
Y_{0}&=& R\sin\varphi \\
Z_{0}&=&l.  \label{2}
\end{eqnarray}
This is the topological invariant of the geometry under study. Here $\theta $
is the polar angle, measured from the axis z, and $\varphi $ the azimuthal
angle.

The coordinates of $P_{1}$, the boundaries of the M\"obius band, are of $%
P_{0}$, the cylinder, plus 
\begin{eqnarray}  \label{3}
X_{1}&=& r\sin\theta \cos\varphi, \\
Y_{1}&=& r\sin\theta \sin\varphi, \\
Z_{1}&=& r\cos\theta. 
\end{eqnarray}
The band has weight $2r$. 

\begin{figure}[]
\centering
\includegraphics[scale=0.4]{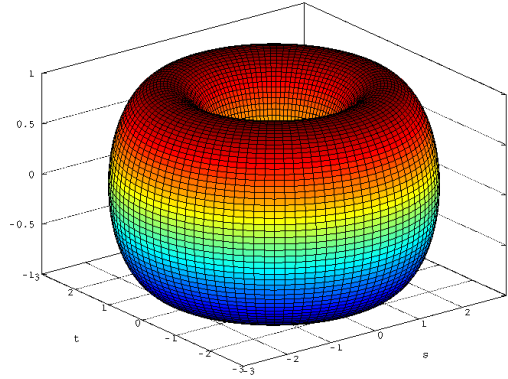}
\caption{(Color online) Torus.}
\label{figtorus}
\end{figure}

This M\"obius strip configuration has a deep relation with the torus and it is a result of 
a reduction that changes both topology and geometry by means of a constraint
\begin{equation}
\theta =\frac{\varphi +\pi }{2}.  \label{5}
\end{equation}
on the Torus geometry ( figure \ref{figtorus}) described by
\begin{equation}
\begin{tabular}{l}
$X=R \cos\varphi +r\ \sin\theta \ \cos\varphi, $ \\ 
$Y=R\ \sin\varphi +r\ \sin\theta \ \sin\varphi, $ \\ 
$Z=l+r\cos\theta. $%
\end{tabular}%
\ \ \ \ \ \   \label{4}
\end{equation}
An important point to enphasize is that the angles are no more independent in the case
of the M\"obius strip, leading to a special embedding from the Torus. In such situation,
the geometry turns to be an unnoriented surface and the topological properties are also altered.

The consideration $r < R$ implies that the generated M\"obius strip lives inside the torus. 
\begin{figure}[]
\centering
\includegraphics[scale=0.4]{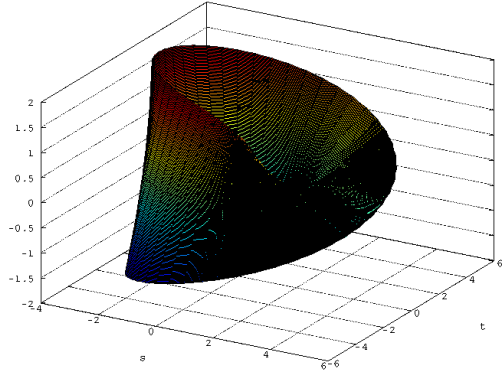}
\caption{(Color online) Mobius.}
\label{figmobius}
\end{figure}
The parametrization associated to the M\"obius strip (figure \ref{figmobius})is then given by
Taking $R=1$ and inserting (5) into (4) we obtain the parametrization of the
band 
\begin{eqnarray}
X &=& \cos\varphi + r \cos\left(\varphi/2\right) \cos\varphi \\
Y &=& \sin\varphi + r\cos\left(\varphi/2\right) \sin\varphi \\
Z &=& l+r\sin\left(\varphi/2\right)  \label{21}
\end{eqnarray}

\section{Lagrange's and Hamilton's of Torus and Mobius}

Taking the parametrization of the Torus we can build the corresponding lagrangean%
\begin{eqnarray}
L_{Torus}&=& \frac{m}{2}\overset{.}{\varphi }^{2}\left[ (R+r\ \sin\theta
\ )^{2}+\frac{r^{2}}{4}\right] \nonumber \\
&+& \frac{m}{2}\left( r\overset{.}{\theta }\right) ^{2}-mr\
\sin\theta \ \overset{.}{Z}_{0}\overset{.}{\theta } \nonumber \\
&+&\frac{m}{2}\left( \overset{.}{Z}_{0}\right) ^{2}
\end{eqnarray}
We can note that by inserting the contraint associated to the reduction to a M\"obius strip, eq. (\ref{5}), the lagrangean
describes the motion on M\"obius strip $L_{Mobius}$.
The Hamiltonian for the torus is easily computed from ($m=R=1$)
\begin{eqnarray}
H_{Torus}=p_{\theta }\overset{.}{\theta }+p_{\varphi }\overset{.}{\varphi }+p_{z_{0}}%
\overset{.}{Z}_{0}-L_{Torus}, 
\end{eqnarray}
where we have
\begin{eqnarray}
p_{\varphi } &=& \frac{\partial L}{\partial \overset{.}{\varphi }}=J_{0},\\
p_{z_{0}} &=& \frac{\partial L}{\partial \overset{.}{Z}_{0}}=-r\ \sin\theta 
\overset{.}{\theta }+\overset{.}{Z}_{0}=L_{0}, \\
p_{\theta} &=& \frac{\partial L}{\partial \overset{.}{\theta }}=r^{2}%
\overset{.}{\theta }-r\ \sin\theta \ \overset{.}{Z}_{0}.  
\end{eqnarray}
This will lead to the following hamiltonian
\begin{align}
H_{Torus}=\frac{1}{2}\left[ \frac{J_{0}^{2}}{(1+r\ \sin\theta \ )^{2}}+\frac{\left(
p_{\theta }+r\sin\theta L_{0}\right) ^{2}}{\left( r\ \cos\theta \ \right) ^{2}}%
+L_{0}^{2}\right],  \nonumber
\end{align} 
by the insertion of eq. (\ref{5}), the torus reduces to a Mobius strip, by Legendre transform $L_{Mobius}$,
we arrive at the above with eq. (\ref{5}), the Hamiltonian for Mobius $H_{Mobius}$.

The quantized version of both hamiltonians then describe 
quantum particles on a Torus and Mobius. Such description has a geometrical base. The deformation of particles turning around on 
such surfaces will not alter the topological structure. The inclusion of the constraint in the Torus, eq. (\ref{5}), 
reduces the geometrically to a Mobius strip inside the Torus. The deformation of the Mobius leads to an intersection, but 
from the topological point of view the topological properties are preserved \cite{diego}.

In order to study the coherent states (CS) associated to the Mobius strip, we analyse again the CS 
torus. As we saw previously, in the torus case, the coordinates $
\theta$ and $\varphi$ are absolutely independent. Thus, we assume two cylinder type 
parametrizations, one for $0\leq l\leq$ $\infty$ cylinder
with angular variable $\varphi$ and other one with finite $0\leq
l_{2}\leq2\pi \sin^{2}\varphi$ $(R=1)$
\begin{equation}
\xi_{Torus}=e^{-\left( l+r\ \cos\theta\right) +i\varphi}\left( 1+r\
\sin\theta\right) e^{-2\pi \sin^{2}\varphi+k\theta} 
\end{equation}
where $i^{2}=k^{2}=-1$. 

The above expression correspond to a geometrical factorization, leading to a
 physical decomposition for $\left\vert
\xi_{Torus}\right\rangle $ that is useful for our proposal is the following
\begin{equation}
\left\vert \xi_{Torus}\right\rangle =\overset{\infty}{\underset{j,m=-\infty }%
{\sum}}\xi_{Mobius}^{-j}e^{-\frac{j^{2}}{2}}\xi^{-m}e^{-\frac{m^{2}}{2}%
}\left\vert j,m\right\rangle,  
\end{equation}
\begin{equation*}
\left\vert \xi_{Mobius}\right\rangle =\overset{\infty}{\underset{j,m=-\infty}{%
\sum}}\xi_{Mobius}^{-j}e^{-\frac{j^{2}}{2}}\left\vert j,0\right\rangle, 
\end{equation*}
i.e., the Mobius strip has its portion splited from the the toroidal space
\begin{align}
\xi_{Mobius} & =e^{-\left( l-r\ \sin\left( \varphi/2\right) \right) +ln\left(
1+r\ \cos\left( \varphi/2\right) \right) +i\varphi}  \tag{63} \\
\xi & =e^{-2\pi \sin^{2}\varphi-r\ \left( \cos\theta+\sin\left( \varphi
/2\right) \right) +ln\left( \frac{1+r\sin\theta}{1+r\ \cos\left(
\varphi/2\right) }\right) +k\theta}, 
\end{align}
this implies in geometrical and topological changes.

\begin{figure}[]
\centering
\includegraphics[scale=0.4]{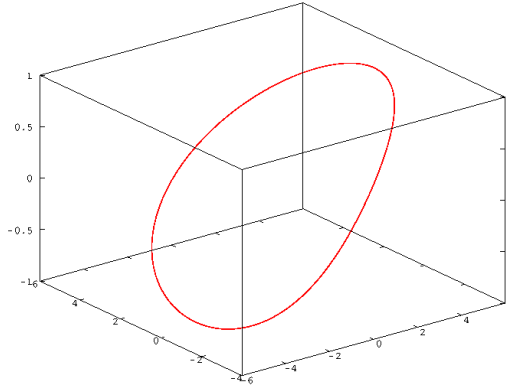}
\caption{(Color online) One particle trajectory on Torus.}
\label{trajtorus}
\end{figure}

\begin{figure}[]
\centering
\includegraphics[scale=0.4]{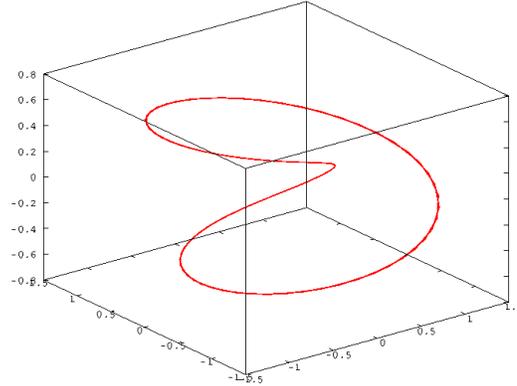}
\caption{(Color online) One particle trajectory on Mobius.}
\label{trajmobius}
\end{figure}

It is interesting to note that, although the topology of the torus is considered as the product of two cilinders, 
the introduction of the parameter $l$ 
involving the $z$ coordinate makes that the Torus becames the product of one cilinder 
with infinity longitude and other with longitude $2\pi$. The link between geometrical description of the Torus and 
its topological equivalence to two cilinders is given by the following equivalent form for  
$\xi_{Torus}$
\begin{eqnarray}
\xi_{Torus} &=& e^{-l +\frac{1}{2}}sgn(\cos\theta) \nonumber \\
&+& i\phi\left(1 + \frac{1}{2} sgn(\sin\theta)\right)e^{-2\pi\sin^{2}\varphi + k\theta},
\end{eqnarray}
With this we can perform the projection from Torus
phase space to the Mobius phase space by means of
\small
\begin{equation}
\left\langle \left\langle \xi _{Mobius}\mid \xi _{Mobius}^{\prime }\right\rangle
\right\rangle =\frac{\left\langle \xi _{Torus}\left\vert \left\vert \xi
_{Mobius}\right\rangle \left\langle \xi _{Mobius}\right\vert \right\vert \xi
_{Torus}^{\prime }\right\rangle }{\left\langle \xi _{Torus}^{0}\left\vert
\xi _{Mobius}\right\rangle \left\langle \xi _{Mobius}\right\vert \xi
_{Torus}^{0}\right\rangle }, \nonumber
\end{equation}
\normalsize
\begin{eqnarray}
=\overset{\infty }{\underset{j=-\infty }{\sum }}%
e^{\left( l^{\prime }+h^{\prime }\right) j}e^{-i\left( \varphi -\psi \right)
}e^{-j^{2}} 
\end{eqnarray}
where $\left\vert \xi _{Torus}^{0}\right\rangle=\left\vert
1_{Torus}\right\rangle =\overset{\infty }{\underset{j,m=-\infty }{\sum }}e^{-%
\frac{m^{2}+j^{2}}{2}}\left\vert j,m\right\rangle $. It is important to
note that we can proceed other time performing the projection from the Mobius 
geometry to the circle straighforwardly obtaining the CS for the Bose
case. Then the procedure of projections can be sinthetized in the following
scheme
\begin{eqnarray}
Torus &\rightarrow& \text{reduction constraint} \nonumber  \\
&\rightarrow&  Mobius \text{ (fermion)} \nonumber \\ 
&\rightarrow& \text{reduction constraint} \nonumber \\
&\rightarrow& Circle \text{ (boson)} \nonumber 
\end{eqnarray}
This in fact connects the aceptions of that the coherent states on Mobius topology as fermions and in the circle topology as
bosons \cite{diego}. A periodic circular trajectory of $2\pi$ on Torus (figure \ref{trajtorus}) 
has a bosonic behaviour that can be deformed topologically 
without losing such characteristic. The Mobius the periodic trajectory of $4\pi$ ( figure \ref{trajmobius}) has a fermionic behaviour 
and can also be deformed without alter the topological properties. Such behaviours in qubits operations acting with deformations that 
preserve topology. 

\section{Entangled states on Mobius}

By considering the coherent states $|\xi\rangle$ on the 
M\"obius strip, we see that their normalization is given in terms of Jacobi theta functions,
as in the case of a torus
\begin{eqnarray}
\langle \xi|\xi\rangle=\Theta_{3}\left(\frac{i}{\pi}\ln|\xi||\frac{i}{\pi}
\right).
\end{eqnarray}
Taking the approximation formula for theta functions, 
\[
\Theta _{3}\left( \frac{i}{\pi }\ln |\xi ||\frac{i}{\pi }\right) \approx
e^{(\ln |\xi |)^{2}}\sqrt{\pi }, 
\]
Then, the the projections relations for coherent
states on a Mobius strip can be written as
\begin{eqnarray}
\langle \xi|\tilde{\xi}\rangle=\sum_{j=-\infty}^{\infty}e^{(l^{\prime }+%
\tilde{l}^{\prime })j}e^{-i(\varphi-\tilde{\varphi})}e^{-j^{2}}, \label{projection}
\end{eqnarray}
where the variables with tilde correspond to the coherent state $|\tilde{\xi}%
\rangle$.

On a period in the Mobius strip, the state turns to the same initial state. In fact, this
property for coherent states defined on a
Mobius strip is a characteristic of the topology associated to this surface. 
Taking $|\xi\rangle=|\xi^{\varphi=0}\rangle$ and $%
|\tilde{\xi}\rangle=|\xi^{\varphi=4\pi}\rangle$, we arrive at 
\begin{eqnarray}
\langle \xi^{\varphi=0}|\xi^{\varphi=4\pi}\rangle&=&\sum_{j=-\infty}^{\infty}e^{2lj + \ln(1+r)j}e^{-j^{2}}\nonumber\\
&=& e^{(2l +\ln(1+r))^{2}/4}\sqrt{\pi} \nonumber \\
&\times&\Theta_{3}\left(-\frac{(2l + \ln(1+r))\pi}{2})|e^{-\pi^{2}}\right). \nonumber \\
\end{eqnarray} 
Using the projection in the form (\ref{projection}), we can write in 
a more simplified form
\begin{eqnarray}
\langle \xi^{\varphi=0}|\xi^{\varphi=4\pi}\rangle = e^{4\pi i} \sum_{j=-\infty}^{\infty}e^{(l^{\prime }+%
\tilde{l}^{\prime })j}e^{-j^{2}}, 
\end{eqnarray}
but as $e^{4\pi i}=1$, we have that the state with a difference in the phase $\varphi$ corresponding to the period
of a Mobius will be the same $|\xi^{\varphi=0}\rangle$, as expected from the topology. Consequently, under a period of
$4\pi$ the state turns to the same. Such behaviour is a characteristic of fermion spin variables and the entanglement 
and in fact can be associated to entanglements for fermionic systems.

Now let us consider the action of the operators restricted to the Mobius strip
\begin{eqnarray}
e^{i\phi_{1}}|j,0\rangle &=& |j+1,0\rangle \\
e^{-i\phi_{1}}|j,0\rangle &=& |j-1,0\rangle \\
T|j,0\rangle&=&|-j,0\rangle.
\end{eqnarray}
As a consequence the action on the state
\begin{equation*}
\left\vert \xi_{Mobius}\right\rangle =\overset{\infty}{\underset{j=-\infty}{%
\sum}}\xi_{Mobius}^{-j}e^{-\frac{j^{2}}{2}}\left\vert j,0\right\rangle, 
\end{equation*}
will be
\begin{equation*}
T\left\vert \xi_{Mobius}\right\rangle =\overset{\infty}{\underset{-j=-\infty}{%
\sum}}\xi_{Mobius}^{j}e^{-\frac{j^{2}}{2}}\left\vert -j,0\right\rangle, 
\end{equation*}
And the state is invariant under the time inversion operator $T$
\begin{eqnarray}
T\left\vert \xi_{Mobius}\right\rangle = \left\vert \xi_{Mobius}\right\rangle. 
\end{eqnarray}
Now let us consider the operators
\begin{equation*}
\hat{M}_{\pm}=\left( e^{\pm i\hat{\phi}_{1}}\otimes T+  T^{-1}\otimes e^{\mp i\hat{\phi}_{1}} \right), 
\end{equation*}%
acting on $|j,0\rangle|j',0\rangle$. This will give
\begin{eqnarray}
\hat{M}_{s}|j,0\rangle|j',0\rangle &=& |j+s,0\rangle|-j',0\rangle + |-j,0\rangle|j'-s,0\rangle, \nonumber
\end{eqnarray}
where $s=\pm 1$.
\begin{figure}[]
\centering
\includegraphics[scale=0.4]{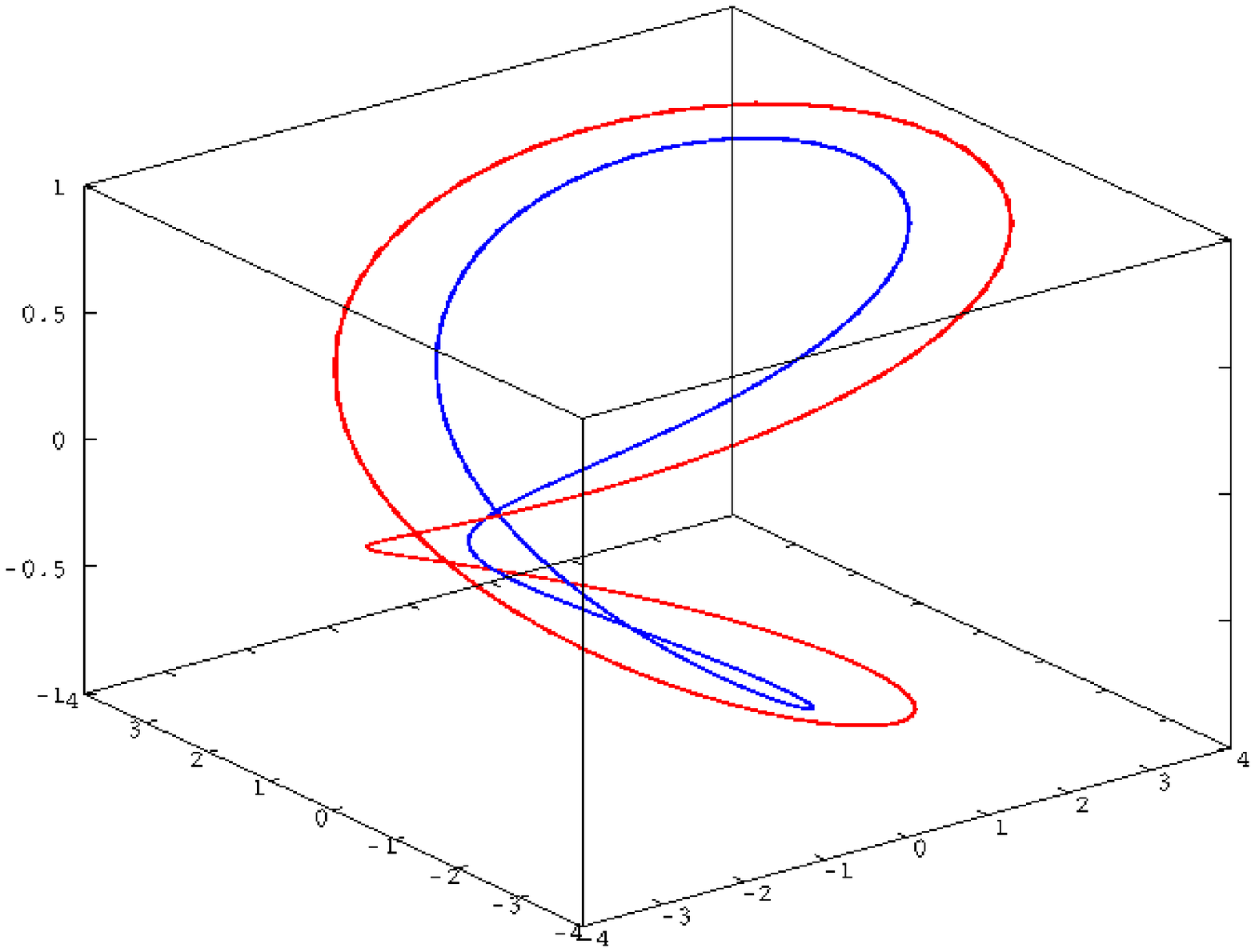}
\caption{(Color online) Two-particle trajectories on Mobius.}
\label{trajtwomobius}
\end{figure}
\begin{figure}[]
\centering
\includegraphics[scale=0.4]{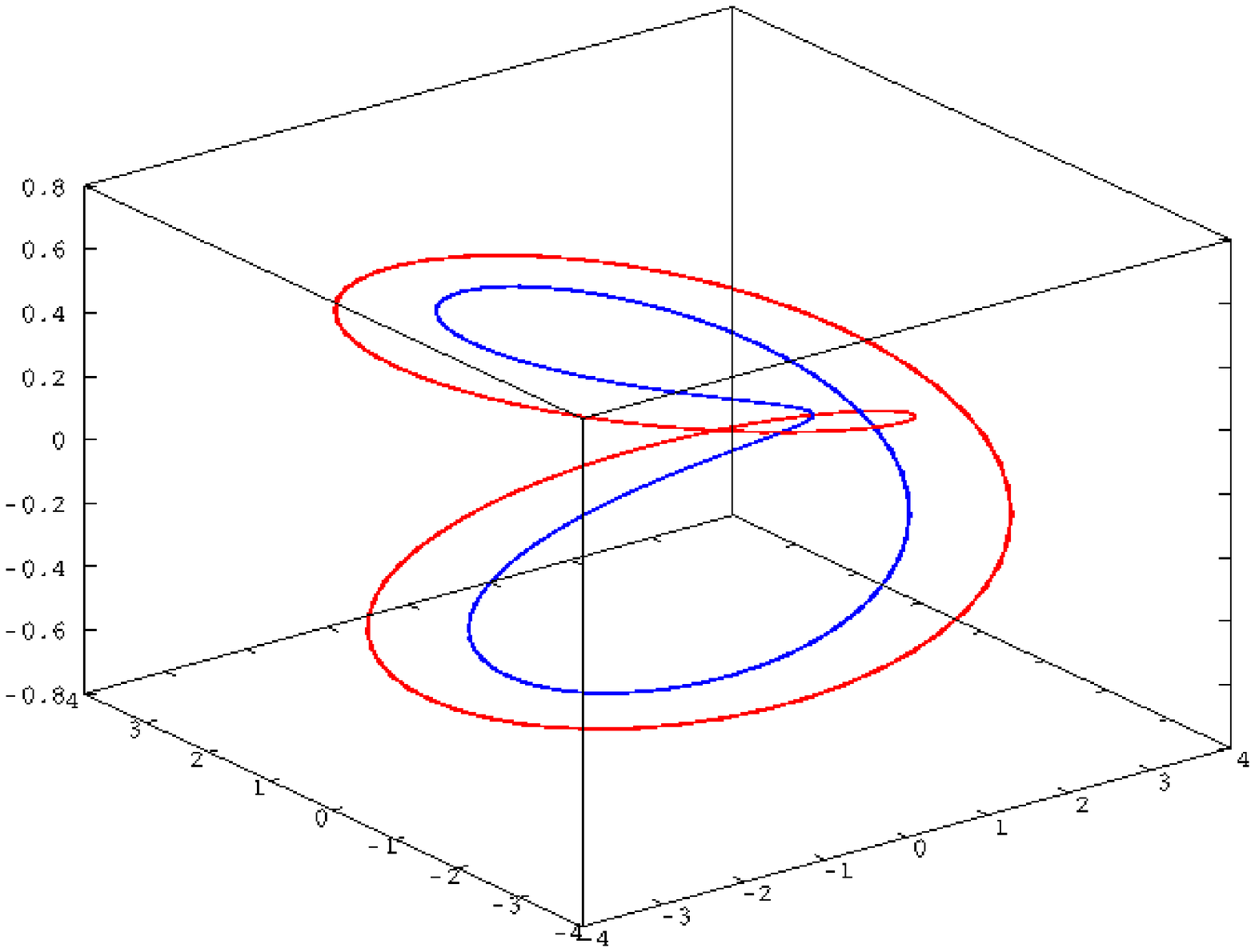}
\caption{(Color online) two-particle trajectories on Mobius.}
\label{trajtwomobius2}
\end{figure}
Let us now consider a two-particle coherent state on the M\"obius strip (figure \ref{trajtwomobius} and \ref{trajtwomobius2}) given by
\begin{eqnarray}
|\psi\rangle = |\xi_{Mobius}\rangle|\tilde{\xi}_{Mobius}\rangle. 
\end{eqnarray}
Under the action of $\hat{M}_{s}$ we will have the entangled state  
\begin{eqnarray}
\hat{M}_{s}|\psi\rangle &=& (e^{is\hat{\phi}_{1}}|\xi\rangle)|\tilde{\xi}\rangle + |\xi\rangle(e^{-is\hat{\phi}_{1}}|\tilde{\xi}\rangle)
\end{eqnarray}
where
\begin{equation*}
e^{is\hat{\phi}_{1}}\left\vert \xi_{Mobius}\right\rangle =\overset{\infty}{\underset{j=-\infty}{%
\sum}}\xi_{Mobius}^{j}e^{-\frac{j^{2}}{2}}\left\vert j+s,0\right\rangle, 
\end{equation*}

In a more general form, we can define
\begin{eqnarray}
\hat{M}=\left( 
\begin{array}{cc}
\hat{M}^{(++)} & \hat{M}^{(+-)} \\ 
\hat{M}^{(-+)} & \hat{M}^{(--)}
\end{array}%
\right),
\end{eqnarray}
that are consistent with the time inversion $T$ and the operator $e^{is\hat{\phi}_{1}}$, where
\begin{eqnarray}
\hat{M}^{(ss')}= \left( e^{ is\hat{\phi}_{1}}\otimes T+  T^{-1}\otimes e^{is'\hat{\phi}_{i}} \right),
\end{eqnarray}
$s,s'=\pm$.
The action on the state $|j,0\rangle|j',0\rangle$ will give
\begin{eqnarray}
\hat{M}^{(ss')}|j,0\rangle|j',0\rangle &=& |j+s,0\rangle|-j',0\rangle + |-j,0\rangle|j'+ s',0\rangle, \nonumber
\end{eqnarray}
and on the state $|\xi\rangle|\tilde{\xi}\rangle$ will give
\begin{eqnarray}
\hat{M}^{(ss')}|\xi\rangle|\tilde{\xi}\rangle &=& (e^{is\hat{\phi}_{1}}|\xi\rangle)|\tilde{\xi}\rangle + |\xi\rangle(e^{is'\hat{\phi}_{1}}|\tilde{\xi}\rangle)
\label{st7}
\end{eqnarray}
We can consider the set of non-orthogonal measurements in terms of coherent states
\begin{eqnarray}
\hat{\gamma}^{(\leftarrow)} &=& \sum_{\xi}|\xi\rangle\langle\xi| \otimes \hat{1}, \nonumber \\
\hat{\gamma}^{(\rightarrow)} &=& \hat{1} \otimes \sum_{\xi}|\xi\rangle\langle\xi|  , \nonumber \\
\langle \xi|\tilde{\xi}\rangle &=& g_{\xi\tilde{\xi}}
\end{eqnarray}
Note that, if we consider the states normalized, this measurement turns a von-Neumann type measurement in each 
reduced space \cite{pors}. By acting on the state (\ref{st7}), we have
\begin{eqnarray}
\hat{\gamma}^{(\rightarrow)}\hat{M}^{(ss')}|\xi\rangle|\tilde{\xi}\rangle &=& (e^{is\hat{\phi}_{1}}|\xi\rangle)\sum_{\xi'}g_{\xi'\tilde{\xi}}|\tilde{\xi}\rangle \nonumber \\
&+& |\xi\rangle \sum_{\xi'}|\xi'\rangle\langle\xi'|(e^{is'\hat{\phi}_{1}}|\tilde{\xi}\rangle)
\end{eqnarray}
\begin{eqnarray}
\hat{\gamma}^{(\leftarrow)}\hat{M}^{(ss')}|\xi\rangle|\tilde{\xi}\rangle &=&  \sum_{\xi'}|\xi'\rangle\langle\xi'|(e^{is\hat{\phi}_{1}}|\xi\rangle)|\tilde{\xi}\rangle \nonumber \\
&+& \sum_{\xi'}g_{\xi'\xi}|\xi\rangle(e^{is'\hat{\phi}_{1}}|\tilde{\xi}\rangle)
\end{eqnarray}

The effective dimensionality of the Hilbert space can be computed in terms of the Hilbert-Schmidt norm
\begin{eqnarray}
Tr(\hat{\gamma}^{(\leftarrow)2})=Tr(\hat{\gamma}^{(\rightarrow)2})
\end{eqnarray}
Consider the density matrix associated to $\hat{M}^{(ss')}|\xi\rangle|\tilde{\xi}\rangle$ is given by $\hat{\rho}_{\xi\xi'}^{ss'}$,
given by
\begin{eqnarray}
\hat{\rho}_{\xi\tilde{\xi}}^{ss'}= \hat{M}^{(ss')}|\xi\rangle|\tilde{\xi}\rangle\langle\tilde{\xi}|\langle\xi|\hat{M}^{(ss')\dagger},
\end{eqnarray}
then the associated entangled measurement can be calculated by means of projective measurements
\begin{eqnarray}
r_{\xi}^{ss'}&=& Tr(\hat{\gamma}^{(\leftarrow)}\hat{\rho}_{\xi\tilde{\xi}}^{ss'}),  \\
r_{\tilde{\xi}}^{ss'}&=& Tr(\hat{\gamma}^{(\rightarrow)}\hat{\rho}_{\xi\tilde{\xi}}^{ss'}), 
\end{eqnarray}
compared to the corresponding measurements to the state $|\xi,\tilde{\xi}\rangle$, that are obtainded by doing $s\rightarrow 0$ 
and $s'\rightarrow 0$,
\begin{eqnarray}
r_{\xi}^{0}&=& Tr(\hat{\gamma}^{(\leftarrow)}\hat{\rho}_{\xi\tilde{\xi}}^{0}),  \\
r_{\tilde{\xi}}^{0}&=& Tr(\hat{\gamma}^{(\rightarrow)}\hat{\rho}_{\xi\tilde{\xi}}^{0}), 
\end{eqnarray}
The ratios
\begin{eqnarray}
\lambda^{ss'}= \frac{r^{ss'}}{r^{0}}
\end{eqnarray}
give the measurement of the entanglement associated do the action of $\hat{M}^{(ss')}$.

\section{Entangled states on Torus-Mobius}

\begin{figure}[]
\centering
\includegraphics[scale=0.4]{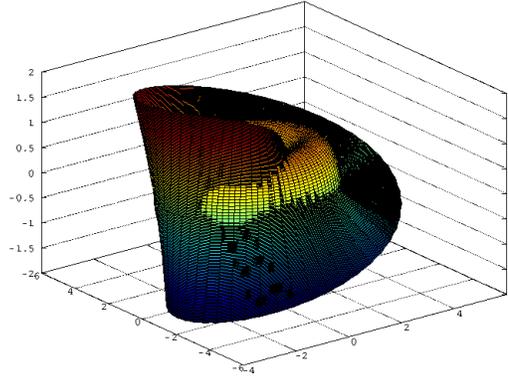}
\caption{(Color online) Intersection of Torus and Mobius.}
\label{mobius_torus}
\end{figure}

\begin{figure}[]
\centering
\includegraphics[scale=0.4]{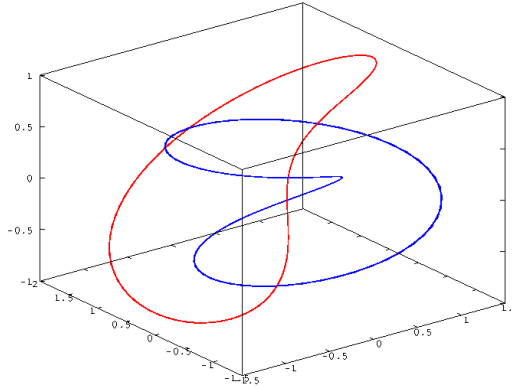}
\caption{(Color online) one-particle trajectories in Torus and Mobius.}
\label{mobius_torusTr}
\end{figure}

We can consider also coherent states pertaining to both spaces, i.e., Torus and Mobius, 
by means of products in the intersection, as
\begin{eqnarray}
|\xi_{Torus}\rangle|\xi_{Mobius}\rangle.
\end{eqnarray}
since we derived the each coherent space separately for Torus and Mobius, we can consider the possibility of 
entanglement between them. Note that this is not an impossible case. In fact, the intersection of the 
Mobius strip and the torus can be a point of interaction between the particles (figure \ref{mobius_torus}).
Although each particle is confined to its surface (figure \ref{mobius_torusTr}), the intersection can be used as
a point of correlation. In fact, if we think in terms of orbital angular momentum and spin, we can associate the interaction 
to a spin-orbit coupling entanglement. 

Let us consider an operator acting on the intersection of Torus and Mobius,
\begin{eqnarray}
\hat{W}_{\bigcap}= \hat{W}^{(Torus)}\otimes\hat{1}^{(Mobius)} + \hat{1}^{(Torus)}\otimes\hat{W}^{(Mobius)} , 
\end{eqnarray}
such that $\hat{1}^{(Mobius)}$ leaves the mobius state invariant and $\hat{1}^{(Torus)}$ leaves the torus state invariant.
The action of $\hat{W}_{\bigcap}$ on $|\xi_{Torus}\rangle|\xi_{Mobius}\rangle$ will be given by
\small
\begin{eqnarray}
\hat{W}^{(Torus)}|\xi_{Torus}\rangle\otimes|\xi_{Mobius}\rangle + |\xi_{Torus}\rangle\otimes\hat{W}^{(Mobius)}|\xi_{Mobius}\rangle \nonumber
\end{eqnarray}
\normalsize
As an example we can calculate $(e^{i\hat{\phi}_{1}})_{\bigcap}$ where
\begin{eqnarray}
(e^{i\hat{\phi}_{1}})_{\bigcap} = e^{i\hat{\phi}_{1}}\otimes\hat{1} + \hat{1}\otimes e^{i\hat{\phi}_{1}} ,
\end{eqnarray}
since the $|j_{1},j_{2}\rangle^{Torus}$ and $|j_{1},0\rangle^{Mobius}$, the action of the operator on these states
will lead to $(e^{i\hat{\phi}_{1}})_{\bigcap}|j_{1},j_{2}\rangle^{Torus}\otimes|j_{1},0\rangle^{Mobius} 
=|j+1,j'\rangle^{Torus}\otimes|j,0\rangle^{Mobius}+|j,j'\rangle^{Torus}\otimes|j+1,0\rangle^{Mobius}$. 
This will have effect in the coherent states by means of 
$(e^{i\hat{\phi}_{1}})_{\bigcap}|\xi_{Torus}\rangle|\xi_{Mobius}\rangle$. A corresponding density matrix 
associated to such state $\hat{\rho}^{Torus-Mobius}$

We can consider the set of non-orthogonal measurements in terms of the coherent states acting on Torus and Mobius
\begin{eqnarray}
\hat{\gamma}^{(Torus)} &=& \sum_{\xi}|\xi\rangle\langle\xi|^{(Torus)} \otimes \hat{1}^{(Mobius)}, \nonumber \\
\hat{\gamma}^{(Mobius)} &=& \hat{1}^{(Torus)} \otimes \sum_{\xi}|\xi\rangle\langle\xi|^{(Mobius)}  , \nonumber 
\end{eqnarray}
\begin{eqnarray}
\langle \xi|\tilde{\xi}\rangle^{(Torus)} &=& g_{\xi\tilde{\xi}}^{(Torus)}, \nonumber \\
\langle \xi|\tilde{\xi}\rangle^{(Mobius)} &=& g_{\xi\tilde{\xi}}^{(Mobius)} \nonumber 
\end{eqnarray}
This measurement turns to a von-Neumann type measurement if we normalize the states. 
The effective dimensionality of the Hilbert space can be computed in terms of the Hilbert-Schmidt norms
\begin{eqnarray}
D^{(Torus)} &=& Tr(\hat{\gamma}^{(Torus)2}) \\
D^{(Mobius)} &=& Tr(\hat{\gamma}^{(Mobius)2})
\end{eqnarray}
The measurement of entanglement can be calculated by means of projective measurements
\begin{eqnarray}
r^{(Torus)} &=& Tr(\hat{\gamma}^{(Torus)}\hat{\rho}^{(Torus-Mobius)}),  \\
r^{(Mobius)} &=& Tr(\hat{\gamma}^{(Mobius)}\hat{\rho}^{(Torus-Mobius)}), 
\end{eqnarray}
compared to the corresponding measurements to the uncorrelated state $\hat{\rho}_{0}$,
\begin{eqnarray}
r_{0}^{(Torus)} &=& Tr(\hat{\gamma}^{(Torus)}\hat{\rho}_{0}),  \\
r_{0}^{(Mobius)} &=& Tr(\hat{\gamma}^{(Mobius)}\hat{\rho}_{0}), 
\end{eqnarray}
by means of the respective ratios
\begin{eqnarray}
\lambda^{(Torus)} &=& \frac{r^{(Torus)}}{r_{0}^{(Torus)}}, \\
\lambda^{(Mobius)} &=& \frac{r^{(Mobius)}}{r_{0}^{(Mobius)}},  
\end{eqnarray}
that give the correlations corresponding to each state on Torus and Mobius.

\section{Conclusion}

We have considered the entanglement of quantum particles in non-trivial topologies, considering 
the cases of a Torus, Mobius strip and Torus-Mobius. We have derived the corresponding parametrizations in each 
case and developed appropriate operators. We have derived the geometrical lagrangeans and 
hamiltonians of Torus and Mobius that are associated to each other by means to a constraint that, from the topological point of view, 
is a topological reductions equivalent to cuts in the deformations. The quantum particle dynamics in the quantized form is a 
consequence the canonical quantization of the hamiltonians. From each case we derived the corresponding one and two-mode coherent 
states that are entangled by the action of the proper operators in each case. We also have shown the relation of the toroidal operators that lead to entanglement and $SU\left( 1,1\right)$, that can lead 
to possible connections with supergravity models.

The Mobius is obtained by means of a reduction from the Torus by means of a constraint in the angular variables.
This implies that the Mobius strip can be keept inside the Torus . By applying a deformation on the strip, 
the topological properties 
keep unaltered and we use it to build the corresponding coherent states associated to the Mobius topology starting from the 
Torus. 

We have shown that the entangled state generated in the Torus has a characteristic of the experimentally verified
 photon entanglement by orbital angular momentum, 
such that the states on Torus behave like bosons as a special case. This is an important fact, since we can associate
the entanglement in a torus to a photon entanglement by orbital angular momentum as a special case. On the other hand, 
the entanglement states associated to Mobius strip 
have the periodicity associated to fermions $4\pi$ and can be more appropriate to describe the entanglement in fermionic 
systems. We also have considered the entanglement between Torus and Mobius in the intersection of Torus and  
Mobius with the action of operators defined in the intersection. Such situation can be originated in the 
case of an entanglement by spin-orbit coupling. By derive non-orthogonal measurements, equivalent to 
the von-Neumann type measurement for orthogonal states, we have shown 
that the generated entangled states can also be evaluated in a consistent way with the appropriate operators in each case. 

The periodic trajectories on Torus ($2\pi$) and Mobius ($4\pi$) surfaces can be deformed topologically 
without alter the typical behaviours. Such behaviours are also reflected 
in the entangled states for Torus and Mobius and are important when we consider deformation operations. We have also shown 
that some operators leave the entangled states invariant, what is important in the case of protecting entanglement.

\section{Acknowledgements}

TP thanks CAPES (Brazil) and
DJCL thanks CNPq (Brazil) for financial support.

\end{document}